\documentclass[aps,prl,nofootinbib,twocolumn,superscriptaddress,preprintnumbers]{revtex4}

\usepackage{amssymb}
\usepackage{amsmath}
\usepackage{epsfig}
\usepackage{hyperref}
\usepackage{breakurl}
\usepackage{xcolor}

\makeatletter
\def\simgt{\mathrel{\lower2.5pt\vbox{\lineskip=0pt\baselineskip=0pt
           \hbox{$>$}\hbox{$\sim$}}}}
\def\simlt{\mathrel{\lower2.5pt\vbox{\lineskip=0pt\baselineskip=0pt
           \hbox{$<$}\hbox{$\sim$}}}}
\makeatother

\def\mysection#1{{\bf #1.} }

\newcommand{\be}{\begin{equation}}
\newcommand{\ee}{\end{equation}}
\newcommand{\bea}{\begin{eqnarray}}
\newcommand{\eea}{\end{eqnarray}}
\newcommand{\beq}{\begin{eqnarray}}
\newcommand{\eeq}{\end{eqnarray}}

\newcommand{\sslash}[1]{\ensuremath\raisebox{-0.00cm}{{\small\slash}}\hspace{-0.21cm}#1\/}

\def\mysection#1{{\bf #1.} }

\def\lsim{\mathrel{\rlap{\lower4pt\hbox{\hskip1pt$\sim$}}
     \raise1pt\hbox{$<$}}}         %less than or approx. symbol
\def\gsim{\mathrel{\rlap{\lower4pt\hbox{\hskip1pt$\sim$}}
     \raise1pt\hbox{$>$}}}         %greater than or approx. symbol

\begin{document}

\widetext
\leftline{MCTP-17-13, PP-017-28} 

\title{Searching for Confining Hidden Valleys at the LHC(b)}

%\affiliation{Michigan Center for Theoretical Physics, University of Michigan, Ann Arbor MI 48109, USA}
\author{Aaron~Pierce}
\affiliation{Michigan Center for Theoretical Physics, University of Michigan, Ann Arbor, MI 48109, USA}
\author{Bibhushan~Shakya}
\affiliation{Department of Physics, University of Cincinnati, Cincinnati, OH 45221, USA}
\affiliation{Michigan Center for Theoretical Physics, University of Michigan, Ann Arbor, MI 48109, USA}
\author{Yuhsin Tsai} 
\affiliation{Maryland Center for Fundamental Physics, Department of Physics,
University of Maryland, College Park, MD 20742, USA}
\author{Yue Zhao}
\affiliation{Michigan Center for Theoretical Physics, University of Michigan, Ann Arbor, MI 48109, USA}

\vskip 0.25cm

%\author{Aaron Pierce, Bibhushan Shakya, Yuhsin Tsai, Yue Zhao}
%\affiliation{Michigan Center for Theoretical Physics, University of
%Michigan, Ann Arbor, MI 48109}
%\affiliation{Department of Physics, University of Cincinnati, Cincinnati, OH 45221, USA}
%\affiliation{Maryland Center for Fundamental Physics, Department of Physics,
%University of Maryland, College Park, MD 20742, USA}

\begin{abstract}
We explore strategies for probing
Hidden Valley scenarios exhibiting confinement. Such scenarios
lead to a multiplicity of light hidden hadrons from showering
processes. Their decays are typically soft and displaced, making them
challenging to probe with traditional LHC searches. We show the
low trigger thresholds and excellent track and vertex reconstruction
at LHCb provide an ideal environment to search for such signals -- in both
muonic and hadronic channels. We also explore the potential of ATLAS/CMS and discuss modifications to
present searches that might make these experiments competitive with
the LHCb reach. Our proposed searches can probe $Z^{\prime}$
models with dominant decays to dark sectors as well as exotic Higgs
boson decays in Twin Higgs models.
\end{abstract}

\maketitle

%%%%%%%%%%%%%\
\section{Motivation}

To ensure no physics remains hidden, experiments at the Large Hadron Collider (LHC) must search for a wide a variety of phenomena beyond the Standard Model (SM). At present, most searches focus on signatures
with energetic particles emerging from a primary vertex or with large missing energy. However, several well-motivated scenarios possess novel decay signatures that are challenging for such traditional LHC searches, motivating new search strategies. We study such strategies for Hidden Valley (HV) scenarios \cite{Strassler:2006im}, which produce soft and displaced objects with small missing energy.

HV scenarios consist of a sector with light (e.g. GeV scale) degrees of freedom connected to the SM sector only via massive particles. This effectively forms a barrier between the sectors. The hidden sector may confine via a non-abelian gauge symmetry \emph{a la} SM QCD, with a model-dependent mass spectrum. An approximate chiral symmetry spontaneously broken by hidden sector confinement leads to light hidden pions, $\pi_v$; otherwise, the
lightest hidden degrees of freedom may be hadrons with masses comparable to or even heavier than the hidden confinement scale $\Lambda_v$, as in e.g. Fraternal Twin Higgs models \cite{Craig:2015pha}. After parton production, showering and hadronization (SH) in the hidden sector produces many hadrons, each much softer than the total energy in the process. Hadron multiplicity can range from a few in the absence of light quarks to ${\mathcal O(1000)}$ \cite{Knapen:2016hky} for a long showering window with large 't Hooft coupling. In this letter we focus on searches for theories averaging ${\mathcal O(10)}$ hadrons per event, and hadron lifetimes long enough to give rise to displaced vertices (DV) at the LHC.  
Complementary strategies that instead use kinematic information in such dark showers are discussed in \cite{Cohen:2017pzm,Cohen:2015toa}.

At LHCb, planned upgrades will remove hardware-level triggers entirely, dramatically improving search flexibility. Furthermore, excellent vertex reconstruction, invariant mass resolution, and particle identification make LHCb an ideal detector for exotic soft long-lived particles. Previous HV searches at LHCb \cite{Aaij:2015ica,Aaij:2016isa} focused on final states with only two hard $\pi_v$ and are not optimal for a larger hadron multiplicity. Our proposed searches instead have similarities with existing LHCb SM searches for $K_S\to\mu^\pm$ \cite{Aaij:2012rt} and $B_0\to D^{\pm}$ \cite{Aaij:2016yip}.

Although stringent trigger requirements and potentially significant backgrounds make HV searches at ATLAS/CMS challenging, these experiments enjoy $\sim20\times$ the luminosity and $\sim 10 \times$ the angular coverage of LHCb. Specialized triggers for displaced decays of long-lived particles exist (e.g. \cite{Aad:2015uaa,ATLAS-CONF-2016-103,CMS:2014wda,Aad:2015rba,ATLAS-CONF-2016-042}), but are limited in efficiency and scope. We study the efficacy of various existing ATLAS/CMS searches for HV scenarios and mention possible modifications to improve their reach.

\section{Setup}

We consider a heavy $U(1)'$ gauge boson $Z_{p}$ with gauge coupling $g_v$, coupled simultaneously to both SM quarks $q$ and HV
quarks $q_{v}$ with $U(1)'$ charges $Q_{v}^{SM,HV}$: \begin{eqnarray}
 \label{HVmodel}
L\supset g_v Z_p^\mu(Q_v^{SM} \bar q\gamma_\mu q+ Q_v^{HV} \bar
q_v\gamma_\mu q_v)+\frac{1}{2}m^2_{Z_p} Z_p^2.
\end{eqnarray}
If $Q_{v}^{SM}\ll Q_{v}^{HV}$, as might be expected if the SM
coupling is induced via kinetic mixing \cite{Holdom:1985ag},
the $Z_p$ primarily decays to the hidden sector; this also alleviates
bounds from direct $Z_p$ searches in leptonic channels
\cite{Curtin:2014cca}. Although the $q_v$ are energetic,
SH processes produce multiple hidden sector hadrons, each substantially less
energetic than the partons. For concreteness, we assume the low
energy HV spectrum possesses two approximately mass degenerate
hadrons: a composite vector boson $\omega_v$ and a pseudo-scalar
$\eta_v$,\footnote{Such a spectrum can be realized in
models with one light quark.} which decay into SM
states. We treat the $\omega_v$ lifetime, which depends on the possible presence of additional mediators, as a free parameter.  We concentrate on scenarios
where the probability to decay within the detector is
non-negligible. Chiral suppression gives the $\eta_v$ a longer lifetime, 
and we assume they escape the detector.\footnote{If the $\omega_v$ lifetime is somewhat shorter than
considered, the signals discussed here might instead be applicable to
$\eta_{v}$ decays.}  The process of interest is:
\begin{eqnarray}\label{eq:zpchannel}
&& q\bar{q}\to Z_p \to q_v\bar{q}_v\to
N_{\omega_{v}}\times\omega_v + N_{\eta_{v}} \times\eta_v,
\nonumber
\\
&&\omega_v\to f\bar{f}\,(\text{displaced decay}).
\end{eqnarray}
We assume $3/4$ of the light mesons are $\omega_v$, as expected from considering the spin degrees of freedom.

Exotic decays of the Higgs boson to HV particles \cite{Strassler:2006ri,Aaij:2016isa} give similar final states:
\begin{eqnarray}\label{eq:THchannel}
&& gg\to h \to q_v\bar{q}_v\to
N_{\omega_{v}}\times\omega_v + N_{\eta_{v}} \times\eta_v.
\end{eqnarray}
We will study this process in a variation of the Twin Higgs (TH) model (see next section).

Showering and hadronization (SH) processes in the hidden sector govern
the average number and kinematics of the hidden hadrons produced. We implement SH processes using a modified version of the Pythia 8.1 Hidden Valley implementation  \cite{Sjostrand:2007gs}, as described in Ref.~\cite{Schwaller:2015gea}.\footnote{We have checked this gives results consistent with the latest Pythia version \cite{Sjostrand:2014zea}, which incorporates running effects.} To test the robustness of our results, we compare with results obtained from a simplified analytical procedure using the Quark-combination model (QCM) \cite{Xie:1988wi, Wang:1996jy, Si:1997zs, Si:1997rp} with the Longitudinal phase space approximation (LPSA) \cite{Webber:1994zd,Han:2007ae}. While different approaches give different average meson multiplicity $\langle
N_v\rangle \equiv \langle N_{\omega_{v}+ {\eta_{v}}} \rangle$, once  $\langle N_v\rangle$ is fixed we find similar kinematic
distributions of mesons. Thus we treat $\langle
N_v\rangle$ as a free parameter.  These different schemes then give comparable results (see
Fig.\,\ref{fig:muonbound}). With this setup, we discuss a variety of search strategies for (i) hidden hadron decay to muon pairs, and (ii) decays to hadrons via heavy quarks, at both LHCb and ATLAS/CMS.

\section {Di-muon Channel}

\subsection{LHCb Strategies}

\begin{figure*}[t!]
\includegraphics[width=8.6cm, height=5.5cm]{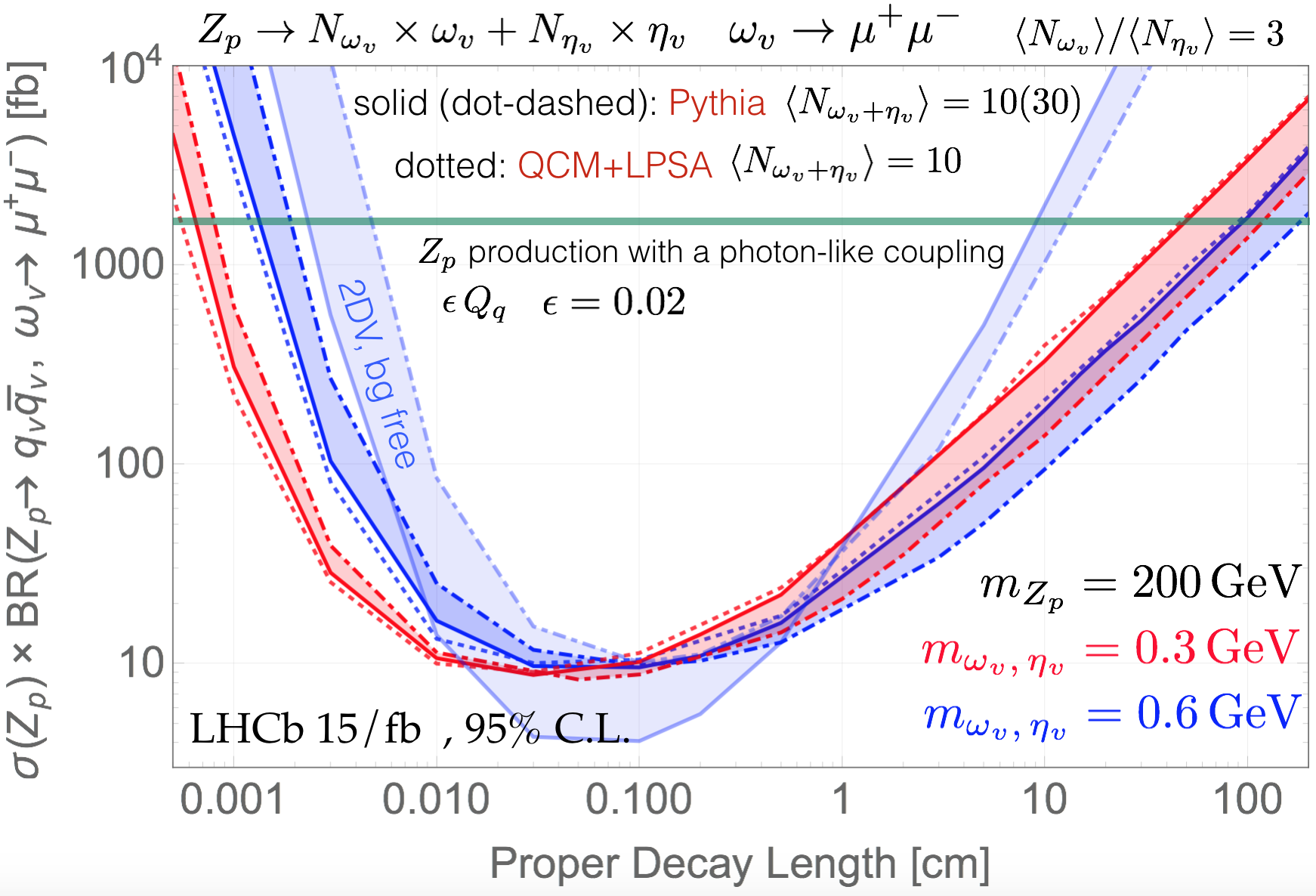}\qquad\includegraphics[width=8.6cm,height=5.5cm]{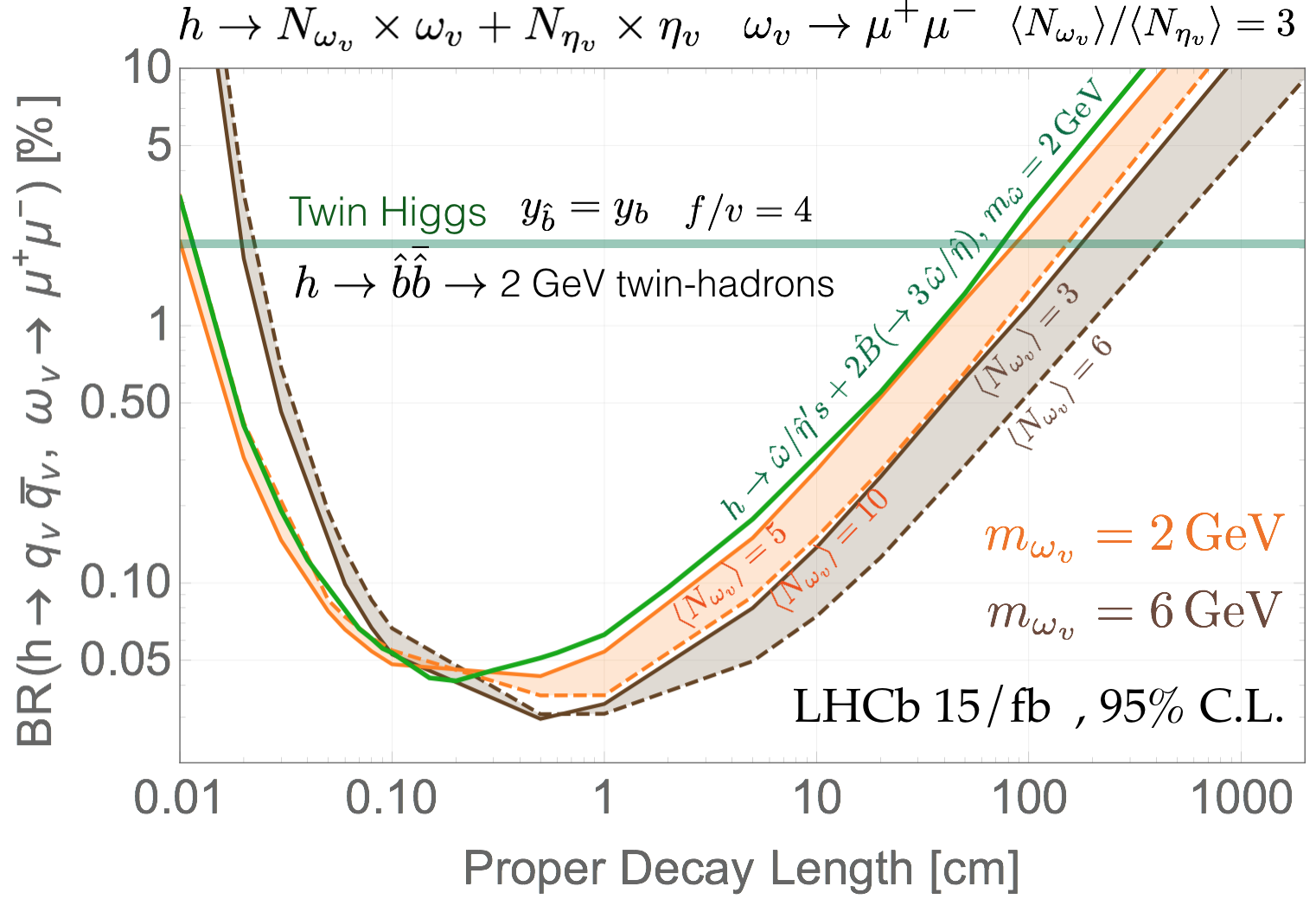}
\caption{Left panel: $Z_p$ cross section reach. 
Green line: cross section for a photon-like coupling, suppressed by $\epsilon=0.02$. Right panel: Projected upper bounds on BR($h\to$ twin bottom quarks) using the 1\,DV search. This process produces lighter twin mesons $\hat{\omega}/\hat{\eta}$ followed by $\hat{\omega}\to\mu^{+}\mu^-$. Horizontal green line: prediction in a variation of the Fraternal Twin Higgs model (see text); in this context $\omega_v$ is a mixture of $c'$ and $s'$.  Green curve: reach for the corresponding decay topology (see text for details).}\label{fig:muonbound}
\end{figure*}

Displaced muon pair searches at LHCb are discussed in \cite{Aaij:2012rt,Ilten:2016tkc}. 
The dark photon search strategy of Ref.\,\cite{Ilten:2016tkc} is directly applicable to our scenario. We apply their selection criteria from the post-module displaced search:\,\footnote{Since we focus on long-lived hidden hadrons, we do not study pre-module event selections.}
\begin{itemize}
\item $\geq\!1$ DV with transverse displacement $\ell_T\in[6,22]$\,mm. 
\item $\eta(\omega_v)\in[2,5]$, each vertex contains two opposite-sign muons with $\eta(\mu^{\pm})\in[2,5]$.
\item $p_T(\mu^{\pm})>0.5$ GeV, $p(\mu^{\pm})>10$ GeV.
\end{itemize}
These requirements ensure that the DV is well-separated from the
beamline and that the two muons can be reconstructed as
tracks in the Vertex Locator (VELO) with muon identification
efficiency $\varepsilon^2_{\mu}\approx0.5$. The transverse
displacement and $\eta$ requirements effectively impose a cut on the longitudinal displacement, forcing the decay to occur within the VELO. 

Fig.\,\ref{fig:muonbound} (left panel) presents LHCb sensitivity to the $Z_p$ model
(Eq.\,(\ref{eq:zpchannel})) with $15$\,fb$^{-1}$ data for $10\leq\langle N_v\rangle\leq 30$, shown as the width of the band. We adopt the background $\approx 25$ events per mass bin from \cite{Ilten:2016tkc}. Results from Pythia and the QCM+LPSA method are shown to illustrate the scheme-independence of our results.  We show results for different hidden hadron masses in red ($m_{\omega_v}$=0.3 GeV) and blue ($m_{\omega_v}$=0.6 GeV).  We also display the reach requiring two DV, assuming a background-free search. For context, we show the cross section of a $Z_p$ with a photon-like coupling suppressed by $\epsilon=0.02$.
 
 Fig.\,\ref{fig:muonbound} (right panel) shows analogous results for exotic decays of the Higgs boson as brown ($m_{\omega_v}$=6 GeV) and orange ($m_{\omega_v}$=2 GeV) bands.  The width of the band corresponds to variation in $\langle N_{\omega_{v}} \rangle$. Here we also explore an incarnation of the Fraternal Twin Higgs model \cite{Chacko:2005pe, Craig:2015pha}, with the hierarchies $\hat{\Lambda}_{QCD}\,\textless\, m_{\hat{c}}\simeq m_{\hat{s}}\,\textless\, m_{\hat{A}}\,\textless\, m_{\hat{b}}\simeq m_{\hat{l},\hat{\nu}}\,\textless\, m_{\hat{t}}$ in the twin sector. We focus on
 decays into twin $b$ quarks, $g g\to h\to
\hat{b}\bar{\hat{b}}$,\footnote{We ignore comparable decays into twin leptons, which
can be invisible or produce similar signals via $\tau'\to\nu'+$hidden hadron.} with subsequent decays $\hat{b}\to \hat{c} \bar{\hat{c}} \hat{s}$.  This eventually produces twin $\hat{\omega}/\hat{\eta}$,
predominantly comprised of the light flavor, which can decay as $\hat{\omega}\to \mu^+ \mu^-$ due to a kinetic mixing
$\epsilon$ between the SM and twin photons.\footnote{Such mixing can be induced by exotic fermions
in UV-complete TH models \cite{Cheng:2015buv, Cheng:2016uqk}. We also
assume the hidden $U(1)^{\prime}$ is broken, and there is no symmetry
preventing the decay of mesons like $\bar c s$.}
The reach is denoted by the green curve.
 Although the light twin mesons are primarily produced via heavy twin
$B$ meson decays (rather than SH processes), the multiplicity is low and the kinematic distributions are similar, resulting in a curve differing only slightly from the orange band.
The prediction in the Twin Higgs scenario outlined above (green line)\footnote{Higgs boson branching
ratios also depend on the ratio of the twin
sector vev $f$ to the SM vev $v$; we take
$f/v=4$, consistent with existing constraints and modest
fine-tuning.} demonstrates LHCb's ability to probe such
models for proper decay length $c\tau_{\hat{\omega}}\lsim 1\,$m,
which can be related to model parameters as
$c\tau_{\hat{\omega}}\simeq
0.1\,\text{m}\left(\frac{\text{GeV}}{m_{\hat{c},\hat{s}}}\right)^3\left(\frac{m_{\hat{A}}}{20\,\text{GeV}}\right)^4\left(\frac{10^{-3}}{\epsilon}\right)^2\left(\frac{\text{GeV}}{\hat{\Lambda}_{QCD}}\right)^2$ \cite{Cheng:2015buv}.
Note, the signal is sensitive to the twin sector spectrum; we postpone studies of such variations to future work.

For  $Z_p$ and Higgs decays, the $\omega_v$ produced are highly boosted. For decay lengths $\gsim 1$ cm, the bounds weaken with increasing lifetime -- decays become less likely to occur in the detector. Below $\lsim\mathcal{O}(0.1)$ cm, the post-module search strategy
is not optimal for signal selection, hence the bounds deteriorate. Likewise, heavier $m_{\omega_v}$ results in smaller boosts; for long (short) lifetime, the decay probability inside the detector is larger (smaller).

For the $Z_p$ case, for longer lifetimes, $c \tau \gsim $ 10 cm, the acceptance for
an individual meson changes by $\textless\,50\%$ for $10\leq\langle N_v\rangle\leq 30$; thus the reach is approximately proportional to $\langle N_v\rangle$. This
behavior can be understood from the softness of the showering
process: the distribution of hadrons from SH peaks at low boost, and this part of the distribution is relatively insensitive to $\langle N_{v} \rangle$.
The high boost portion has more variation with
$\langle N_{v} \rangle$; however, the requirement that decays occur inside
the VELO removes precisely this portion.  
For sufficiently short decay lengths, even highly boosted particles decay within the detector, resulting in greater sensitivity to $\langle N_{v} \rangle$, with stronger bounds for smaller $\langle N_v\rangle$.

\subsection {ATLAS\,/\,CMS searches}

We consider three existing triggers.  The first, multiple displaced soft muons, is naively the closest fit to our topology.   The other two, $\sslash{E}_T$ and a hard displaced muon, can give meaningful bounds despite not exactly matching the signal features. Ordinary multi-lepton searches reject muons with large displacements to remove cosmic-ray muon background \cite{ATLAS-CONF-2016-075} and do not apply.  Searches for displaced objects relax such requirements, thus this background can dominate and requires proper treatment.

\begin{figure}
\includegraphics[width=8.7cm,height=5.6cm]{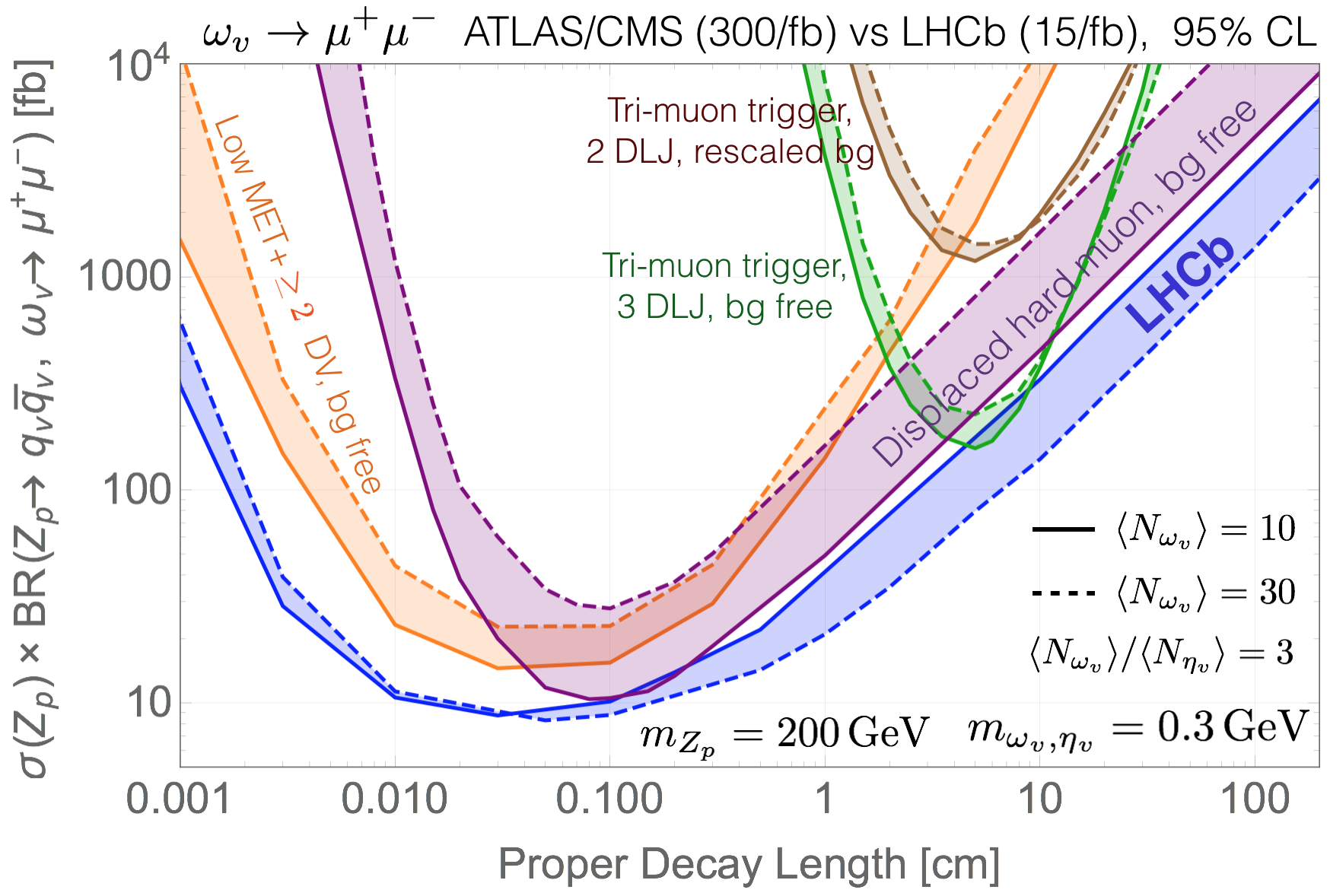}
\caption{Projected bounds from various ATLAS/CMS displaced muons
search strategies, see text for details. The brown curve represents an extrapolation of a current analysis, while the green curve represents only a minor modification.  The orange and purple projections have aggressive assumptions about backgrounds and will likely weaken following detailed detector simulations. The band widths correspond to $10\leq\langle N_v\rangle\leq 30$. The blue band is derived from the LHCb search proposed in this work.}
\label{fig:LHCbound}
\end{figure}

\textit{(i) Multiple displaced soft muons.}
Ref.\,\cite{ATLAS-CONF-2016-042} used a tri-muon trigger for
displaced lepton jet (DLJ) searches, selecting events with at least three
muon spectrometer (MS)\,-\,only tracks with $p_T\,\textgreater\, 6$ GeV
within a $\Delta R<0.4$ cone.  In a HV scenario, this requires at
least two hidden hadrons to decay between $4\,\text{m}\,\textless\,
r\,\textless\,6.5\,$m within $\Delta R\,\textless\,0.4$.  The
analysis required an additional displaced decay with $\Delta R\geq 0.4$ separation from
the ones used for the trigger, with a more inclusive $10\,\text{cm}\,\textless\,
r\,\textless\,6.5\,$m. This necessitates the decay of a third
hidden hadron.
We require $p_T\,\textgreater\,6$ GeV for each muon.
For the reconstruction efficiency of a muon pair from hidden hadron decay, we adopt the
value of $0.4$ reported in the analogous CMS displaced muon
search \cite{CMS:2014hka} for the decay of a $150$ GeV particle,
although a lower efficiency may be expected for softer muons. To estimate background, which may be
dominantly from cosmic muons, we rescale the background from
\cite{ATLAS-CONF-2016-042} to $300$ fb$^{-1}$, optimistically
assuming systematic uncertainties improve as $\sqrt{N}$.
The corresponding reach curve (Fig.\,\ref{fig:LHCbound}, 2 DLJ -- brown band) is worse than that from LHCb (blue band, corresponding to the red band of Fig.~\ref{fig:muonbound}).
One possibility to reduce the background is to require yet another displaced vertex with the looser $r$ cut and $\Delta R\geq 0.4$ from the initial DLJs. For lifetimes such that displaced hadron decays within the detector are likely, this search 
(3 DLJ -- green band) can be competitive with LHCb assuming negligible
background.  Requiring too many hidden hadrons to
decay within a small region of the detector hurts signal acceptance. Relaxing this by,
e.g., including the inner detector for the tri-muon trigger could improve the reach dramatically.

\textit{(ii) $\sslash{E}_T$ trigger}. For long lifetimes, most hidden hadrons escape the detector without decaying and contribute to missing energy,\footnote{Here we define $\sslash{E}_T$ as the opposite of the vector sum of all well-constructed objects. We do not include the (displaced) muons in this category.
With this choice, our $\sslash{E}_T$ definition should align with that in
\cite{ATL-PHYS-PUB-2015-023,Khachatryan:2014gga}.} motivating a
search for $\sslash{E}_T$ in combination with soft (displaced) muons. An ATLAS $\sslash{E}_T$+DV search \cite{Aad:2015rba} triggered on events with $\sslash{E}_T$,  then further (at the analysis level) required a hard muon ($p_T>55$ GeV). To optimize for our signal, we relax the $\sslash{E}_T$ down to the trigger requirement \cite{Sirunyan:2017hci}.  We then relax the muon $p_T$ cut, demanding instead two dimuon DVs:\,\footnote{The idea of using multiple displaced vertices in similar
scenarios was also discussed in Ref.\,\cite{Cui:2014twa}.}
\begin{itemize}
\item $\sslash{E}_T\,\geq 110$ GeV.
\item Reconstruct $\geq 2$ DVs, each with $\ell_T\in[1,30]$ cm.
\item Each muon has $p_T>10$ GeV.
\end{itemize}
The DV $\ell_T$ requirement ensures they are inside the silicon
detector (before the transition radiation tracker) in order to
reconstruct the muon track in the tracker. With a displaced vertex reconstruction efficiency of 0.4 and an optimistic assumption that the search is background-free, the corresponding reach is shown in Fig.\,\ref{fig:LHCbound} (orange
band), which is weaker than our LHCb projection.

\textit{(iii) Hard displaced muon.} Following the analysis in \cite{Aad:2015rba}, we require:
\begin{itemize}
\item  $\geq 1$ reconstructed DV, each with $\ell_T\in[1,30]$ cm.
\item $\geq 1$ muon with $p_T\geq 55$ GeV and transverse impact factor $>1.5$ mm, and $p_T\,\textgreater\, 10$ GeV for the other displaced muon from this vertex.
\end{itemize}
For light hidden hadrons, the DV invariant mass cut imposed in \cite{Aad:2015rba} must be removed. There are already $\mathcal{O}(1)$ background events in the low invariant mass region in this 8 TeV, 20.3/fb search (see
Fig.\,9 in Ref.\,\cite{Aad:2015rba}); the precise number is
difficult to estimate since the dominant background is likely
combinatoric in nature. Regardless, this search has a slightly weaker projected reach than LHCb even when assumed to be background-free (Fig.\,\ref{fig:LHCbound}, purple band). We expect the sensitivity to significantly improve if the stringent $p_T(\mu)$ cut, detrimental for the generically soft hidden hadrons from SH
processes, could be relaxed while maintaining low background.

\section{Heavy Flavor Decay Channels}

We focus on the $c\bar c$ channel and comment on the $b \bar b$ channel later.

\subsection{LHCb Strategies}
For hidden hadron decays to  $c\bar c$, subsequent SM hadronization often produces two $D$ mesons, $D_{(s)}^0$,
$\bar D_{(s)}^0$ or $D^\pm$. The non-negligible lifetimes of charmed hadrons creates an additional separation between the DV from the two $D$ meson decays, resulting in two vertices with large separation from the primary vertex and a small but significant separation from each other. Resolving the secondary vertices should be straightforward as the position resolution in the VELO is $\mathcal{O}(10)\mu$m while a $D$ meson has proper decay length $\mathcal{O}(100)\mu$m. We do not explicitly impose a minimum separation requirement between the two vertices, but note this could be implemented as an additional powerful handle to reject background if necessary. Several strategies to identify $D$ mesons at LHCb exist; some do not require the $D$ meson to be fully reconstructed \cite{Aaij:2015yqa}. Given the sizable probability for a single $D$ meson decay to create three or more charged tracks that can be well-reconstructed in the VELO,
we consider the following increasingly inclusive search strategies:
\begin{itemize}
\item{Two nearby reconstructed displaced $D$ mesons. (2D)}
\item{One displaced reconstructed $D$ meson and one DV with $\geq 3$ charged tracks nearby. (1D1V)}
\item{Two DV, each with $\geq 3$ charged tracks, near each other. (2V)}
\end{itemize}
When reconstructing a $D$ meson, we focus on decay channels containing only charged particles: $BR(D^+\to
K^-2\pi^+)\sim 9.5\%$, $BR(D^0\to K^-2\pi^+\pi^-)\sim8\%$, and
$BR(D_s^+\to K^+ K^-\pi^+)\sim5\%$. Meanwhile, requiring $\geq 3$ charged tracks in a DV rather than a reconstructed $D$ meson selects the decays $BR(D^+\to
3-\textrm{prong})\sim 18.6\%$, $BR(D^0\to 4-\textrm{prong})\sim
14.5\%$ and $BR(D_s^+\to 3-\textrm{prong})\sim 12\%$.

A well-defined track is required to have $p_T\,\textgreater\,0.1$ GeV
and hit at least three disks in the VELO; with these, we expect good track reconstruction resolution
despite the displacement from the primary vertex. To retain $\sim90\%$ of the signal while rejecting cosmic rays and combinatoric background, we impose varying {\it maximum} displacement requirements. For 2D events, the boost factor for each $D$ meson is known, and we require $d_D\lsim5 c\tau_D\gamma_{max}$, where $d_D$ is the distance between the two $D$ mesons in the lab frame, and $\gamma_{max}$ is the larger of the two $D$ meson boost
factors. For 1D1V events, we require $d_D< 5 c\tau_D\gamma_D$ and $d_{T,D}<8 c\tau_D$,
where $d_{T,D}$ is the transverse distance between the $D$ meson vertex and the DV. Finally, for 2V events, we require $d_{T,D}< 5 c\tau_D$ and $d_{Z,D}< 30 c\tau_D$, where $d_{Z,D}$ is the distance between the two vertices along the beam direction. We require that the $D$ meson decay $>10c\tau_D$ away from the primary vertex to eliminate SM heavy quark backgrounds.  Following the $B^0\to D^+D^-$ search in \cite{Aaij:2016yip}, when reconstructing a $D$ meson, we require the
scalar sum of the $p_T$ of the charged tracks exceed $1.8$ GeV. When identifying a $D$ meson pair, we require the scalar sum of the $p_T$ of the two mesons and the total momentum exceed $5$ and $10$ GeV respectively.

\begin{figure}
\includegraphics[width=8.7cm, height=5.6cm]{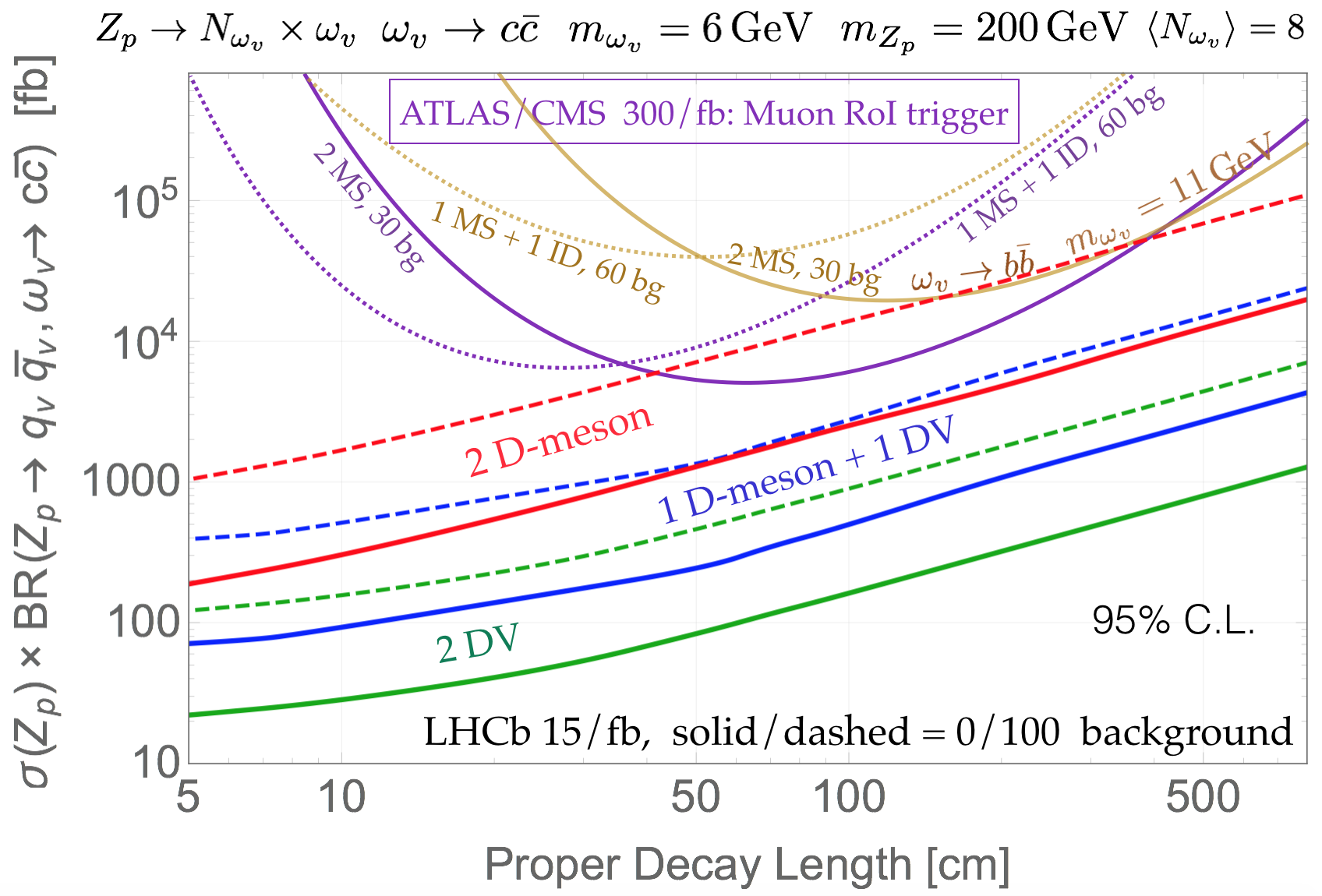}
\caption{Projected bounds from various displaced $c\bar{c}$ search strategies, see text. Purple curves: ATLAS/CMS reach estimate for DV decays into $\geq 5$ charged tracks, with either two DV in the muon spectrometer (solid) or one DV in the inner detector and one in the muon spectrometer (dotted). 
Brown: analogous ATLAS/CMS reach for $\omega_v\to b\bar{b}, m_{\omega_v}=11$ GeV. }\label{fig:Dmeson}
\end{figure}

The background is expected to come dominantly from combinatorics
\cite{Aaij:2016yip}, necessitating a detailed detector simulation. Since this is beyond the
scope of this paper, we instead present our results in
Fig.\,\ref{fig:Dmeson} assuming $0,100$ background events with 15 $\textrm{fb}^{-1}$ data.
 The limits degrade with
increasing lifetime because the decay particles are
more likely to escape the VELO without leaving a sufficient number
of hits. The limits also degrade at shorter decay lengths (analogous
to Figures\,\ref{fig:muonbound},\,\ref{fig:LHCbound}), but we do not
plot this region as SM backgrounds are expected to become
important for displacement $\lsim\mathcal{O}\,$(cm).

\subsection {ATLAS\,/\,CMS searches}

An ATLAS search for long-lived hadronic decays
\cite{Aad:2015uaa}, targeting HV $Z_{p}$ and stealth
supersymmetry models, 
applies two different triggers: jet+$\sslash{E}_T$, and a Muon
Region of Interest (RoI) cluster trigger. Post trigger, both require $\geq 2$\,DV at the analysis level. Two types of DV are considered: an inner detector (ID)
vertex, covering $5\,\text{cm}\,\textless\, r \,\textless\, 30\,\text{cm}$, or an MS
vertex, covering the region between the outer edge of the
hadronic calorimeter and the middle station of the muon chambers, $4 \, \rm{m}\,\textless\, r\,\textless\, 6.5\,\rm{m}$. The jet+$\sslash{E}_T$ trigger requires a leading jet
with $p_T\,\textgreater\, 120$ GeV, $\sslash{E}_T\textgreater$\,200 GeV, and $\geq 7$ charged tracks in each DV.
Given the small probability of getting $\geq 7$ tracks in our scenarios (see BR into 3--prong and 4--prong topologies listed earlier), this trigger is not optimal for our signal.

In the Muon RoI cluster triggered events, $\geq 5$ charged tracks are required at each DV. Triggering the signal requires $\geq\,3$ tracks with $p_T>10$ GeV from the MS vertices.  We estimate bounds for both two MS displaced decays or one MS vertex and one ID vertex in Fig.~\ref{fig:Dmeson}, rescaling the background in \cite{Aad:2015uaa} to 300 fb$^{-1}$. While this extension naively appears weaker than the LHCb reach, further optimization might yield improvement. For instance, the number of charged tracks required in Ref.\,\cite{Aad:2015uaa} was optimized for specific benchmark models and may not be ideal for all scenarios. Requiring more displaced vertices, however, does not appear promising: while this may eliminate background, we find that the signal efficiency is such that the reach worsens.

The above analyses can be repeated without major modifications for the $b\bar b$ channel as the
decay of a $B$ meson is very likely to produce a $D$ meson.
At LHCb, more detailed $B$ meson reconstruction will likely allow further
background rejection, leading to comparable or potentially better reach than the $c\bar c$ channel. At ATLAS/CMS, a greater number of tracks in $b\bar b$ decay favors the $\geq 7$ charged tracks requirement but reduces the $p_T$ of each track, resulting in worse reach than the $c\bar c$ channel (Fig.\,\ref{fig:Dmeson}, brown curves). The relative sensitivity to  the $c\bar c$ and $b\bar b$ channels depends on $m_{Z_p}$ as well as $\langle N_{\omega_v} \rangle$.

%%%%%%%%%%%%%%%%%%%

\section{Conclusions}

We have demonstrated that LHCb provides an ideal environment to search for decays of soft long-lived particles in dimuon and heavy flavor channels as realized in confining Hidden Valley scenarios, with good reach for $Z_p$ models and a realization of the Twin Higgs model. 
Projections for current ATLAS/CMS strategies appear comparatively worse, as even the displaced searches often require energetic final state particles or large missing energy, which are inefficient for HV searches. Our studies indicate that using additional displaced vertices in the trigger or as a background discriminant in place of such stringent cuts could lead to dramatic improvements in sensitivity, potentially making ATLAS/CMS competitive with LHCb. Related ideas have recently been discussed in Ref.\,\cite{Gershtein:2017tsv}.

%%%%%%%%%%%%%%%%%%%
~\\
\mysection{Acknowledgments}
We are grateful to David Curtin, Yuanning Gao, Brian Hamilton,
Philip Ilten, Hassan Jawahery, Simon Knapen, Daniel Levin, William
Parker, Heather Russell, Raman Sundrum, Jesse Thaler, Andreas
Weiler, Mike Williams and Zhenwei Yang for useful discussions. This
work was performed in part at the Aspen Center for Physics, which is
supported by National Science Foundation grant PHY-1066293. AP and
YZ are supported by US Department of Energy under grant DE
-SC0007859. YT is supported by the NSF under grant NSF-PHY-1620074
and the Maryland Center for Fundamental Physics. BS acknowledges
support from the University of Cincinnati and thanks the CERN and
DESY theory groups, where part of this work was conducted, for
hospitality.

\bibliography{HVbib}

%\begin{thebibliography}{99}
%\end{thebibliography}

\end{document}